# EXISTENCE AND STATIC STABILITY OF A CAPILLARY FREE SURFACE APPEARING IN A DEWETTED BRIDGMAN PROCESS.I.

## Agneta M. Balint[1a], Stefan Balint[2b]

[1]*Department of Physics, West University of Timisoara, Bulv. V. Parvan 4, 300223 Timisoara, Romania*

[2]*Department of Computer Science, West University of Timisoara, Bulv. V. Parvan 4, 300223 Timisoara, Romania*

[a)]Corresponding author: agneta.balint@e-uvt.ro

**Abstract.** This paper present six theoretical results concerning the existence and static stability of a capillary free surface appearing in a dewetted Bridgman crystal growth technique. The results are obtained in an axis symmetric 2D model for semiconductors for which $\theta_c + \alpha_e < \pi$ (where: $\theta_c$ - wetting angle and $\alpha_e$ - growth angle). Numerical results are presented in case of InSb semiconductor growth. The reported results can help, the practical crystal growers, in better understanding the dependence of the free surface shape and size on the pressure difference across the free surface and prepare the appropriate seed size, and thermal conditions before seeding the growth process.

## INTRODUCTION

Dewetted Bridgman is a crystal growth technique based on the Bridgman method in which the crystal is grown detached from the ampoule wall by the free surface of a liquid bridge at the level of the liquid-solid interface. The liquid bridge is called meniscus. (see Fig.1)

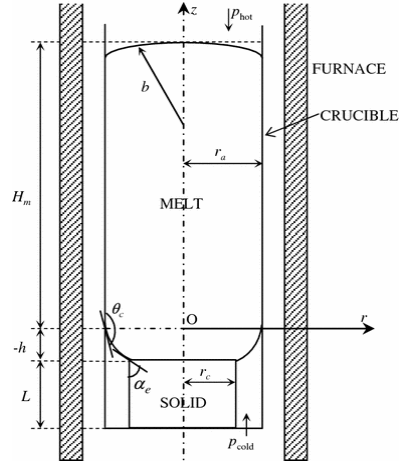

Fig1 Schematic dewetted Bridgman crystal growth system.

Dewetting was first obtained spontaneously in space experiments during InSb Bridgman solidification performed on Skylab-NASA mission-1974 [1],[2] and subsequently in many experiments carried out in orbiting spacecraft's on a wide variety of semiconductors [3]. Understanding the results obtained in microgravity opened the possibility for the dewetting growth on the Earth that can be obtained by applying a gas pressure between the cold and hot sides of the sample $\Delta P = P_{cold} - P_{hot}$ [4],[5] (Fig1.).An experimental application of this is described in [4].On the basis of results reported by Duffar the conditions of detached solidification under controlled pressure difference were investigated by Palosz et.al in [6].Using un-coated and coated-silica ampules they achieved detached and partially detached growth in some 20 solidification experiments. They concluded: if $\theta_c + \alpha_e > \pi$ ,then steady detached growth is possible in a wide range of pressure differences; if $\theta_c + \alpha_e < \pi$, then a steady state detached growth may be expected in a narrow range only.  In [7] the dependence of the meniscus shape on the pressure difference were analyzed for $\theta_c + \alpha_e < \pi$. This paper intends to improve the theoretical part of the analysis presented in [7] taking into account also on the static stability condition of the meniscus.  The analysis is developed in an axis symmetric 2D model. The differential equation of the meridian curve of the meniscus free surface in the coordinate system presented in Fig.1 is given by:

$$z'' = \frac{-\rho \cdot g \cdot z + p}{\gamma}\left[1+(z')^2\right]^{3/2} - \frac{1}{r}\cdot\left[1+(z')^2\right]\cdot z'; \quad R_c \leq r \leq R_a \quad (1)$$

where $R_c$ is the crystal (seed) radius, $R_a$ is the ampule radius and $p$ is the pressure difference $p = \rho \cdot g \cdot H_m - \Delta P$. The function $z(r)$ satisfying (1), has to verify also the following conditions:

$$z'(R_c) = \tan((\pi/2) - \alpha_e) \; ; \qquad z(R_c) = -h_c \quad (2)$$

$$z'(R_a) = \tan(\theta_c - \frac{\pi}{2}) \quad ; \qquad z(R_a) = 0 \quad (3)$$

$z(r)$ is strictly increasing on $[R_c, R_a]$ (4)

Beside the conditions (1) - (4) the function $z(r)$ ,describing the meridian curve, has to minimize the energy functional of the melt column behind the free surface. This functional is given by:

$$I(z) = \int_{R_c}^{R_a} \left\{ \gamma \cdot \left[1+(z')^2\right]^{1/2} - \frac{1}{2}\cdot\rho\cdot g \cdot z^2 + p \cdot z \right\} \cdot r \cdot dr \quad (5)$$

The last condition is called the static stability condition of the axis symmetric free surface. It is essential because in real world equilibrium capillary free surfaces exist only when the minimum condition is satisfied [8].

# THEORETICAL RESULTS.

**Statement 1.** If $\theta_c + \alpha_e < \pi$, then a necessary condition for the existence of a function $z(r)$ having the properties (1) - (4) and $z''(r) < 0$ for $r \in [R_c, R_a]$ (i.e. concave meridian curve) is that the pressure difference $p = \rho \cdot g \cdot H_m - \Delta P$ verifies the inequalities:

$$l(\varepsilon) = \gamma \cdot \frac{\theta_c + \alpha_e - \pi}{\varepsilon} \cdot \sin\theta_c - \rho \cdot g \cdot \varepsilon \cdot \tan((\pi/2) - \alpha_e) - \frac{\gamma}{R_a} \cdot \cos\theta_c \leq \quad (6)$$

$$p \leq \gamma \cdot \frac{\theta_c + \alpha_e - \pi}{\varepsilon} \cdot \sin\alpha_e + \frac{\gamma}{R_a - \varepsilon} \cdot \cos\alpha_e = L(\varepsilon)$$

Here $\varepsilon = R_a - R_c$ = the size of the gap between the crystal(seed) and ampule walls.

**Statement 2.** If $\theta_c + \alpha_e < \pi$, and $0 < \varepsilon' < R_a$ then a sufficient condition for the existence of a number $\varepsilon$ verifying $0 < \varepsilon \leq \varepsilon'$ and a function $z(r)$ having the properties (1) - (4) and $z''(r) < 0$ for $r \in [R_c, R_a]$ (i.e. concave meridian curve) is that the pressure difference $p = \rho \cdot g \cdot H_m - \Delta P$ verifies the inequality:

$$p < \gamma \cdot \frac{\theta_c + \alpha_e - \pi}{\varepsilon'} \cdot \sin\theta_c - \rho \cdot g \cdot \varepsilon' \cdot \tan\left(\frac{\pi}{2} - \alpha_e\right) - \frac{\gamma}{R_a} \cdot \cos\theta_c = l(\varepsilon') \quad (7)$$

Here $\varepsilon = R_a - R_c$ = the size of the gap between the crystal(seed) and ampule walls.

**Statement 3.** If $\theta_c + \alpha_e < \pi$, and $0 < \varepsilon_1 < \varepsilon_2 < R_a$ then a sufficient condition for the existence of a number $\varepsilon$ verifying $\varepsilon_1 < \varepsilon \leq \varepsilon_2$ and function $z(r)$ having the properties (1) - (4) and $z''(r) < 0$ for $r \in [R_c, R_a]$ (i.e. concave meridian curve) is that:

$$L(\varepsilon_1) = \gamma \cdot \frac{\theta_c + \alpha_e - \pi}{\varepsilon_1} \cdot \sin\alpha_e + \frac{\gamma}{R_a - \varepsilon_1} \cdot \cos\alpha_e <$$

$$< \gamma \cdot \frac{\theta_c + \alpha_e - \pi}{\varepsilon_2} \cdot \sin\theta_c - \rho \cdot g \cdot \varepsilon_2 \cdot \tan\left(\frac{\pi}{2} - \alpha_e\right) - \frac{\gamma}{R_a} \cdot \cos\theta_c = l(\varepsilon_2) \quad (8)$$

and the pressure difference $p = \rho \cdot g \cdot H_m - \Delta P$ verifies the inequalities:

$$L(\varepsilon_1) < p < l(\varepsilon_2) \quad (9)$$

Here $\varepsilon = R_a - R_c$ = the size of the gap between the crystal(seed) and ampule walls.

**Statement 4.** A sufficient condition of static stability /instability of the 2D axis symmetric capillary free surface of the meniscus which meridian curve is the function $z(r)$ having the properties (1) - (4) and $z''(r) < 0$ for $r \in [R_c, R_a]$ (i.e. concave meridian curve) is that the inequalities:

$$\frac{\varepsilon}{(R_a - \varepsilon)^{1/2}} < \pi \cdot \frac{1}{R_a^{1/2}} \cdot \frac{\gamma^{1/2} \cdot \sin^{3/2}\alpha_e}{\rho^{1/2} \cdot g^{1/2}} \quad \text{and} \quad \frac{\varepsilon}{(R_a - \varepsilon)^{1/2}} > 2\pi \cdot \frac{1}{R_a^{1/2}} \cdot \frac{\gamma^{1/2} \cdot \sin^{3/2}\theta_c}{\rho^{1/2} \cdot g^{1/2}} \quad (10)$$

hold, respectively.

The next statement is a necessary condition concerning the pressure difference for which a convex-concave (i.e. $z''(R_c) > 0$ and $z''(R_a) < 0$) meniscus having a gap size $\varepsilon$ exist.

**Statement 5.** If $\theta_c + \alpha_e < \pi$ then a necessary condition for the existence of a function $z(r)$ having the properties(1)-(4) and $z''(R_c) > 0$, $z''(R_a) < 0$ is that the pressure difference $p = \rho \cdot g \cdot H_m - \Delta P$ verifies inequalities:

$$l(\varepsilon) = -\frac{\gamma}{R_a - \varepsilon} \cdot \cos\alpha_e - \rho \cdot g \cdot \varepsilon \cdot \tan(\frac{\pi}{2} - \alpha_e) < p < -\frac{\gamma}{R_a} \cdot \cos\theta_c = L \qquad (10)$$

Here $\varepsilon = R_a - R_c$.

**Statement 6**. If $\theta_c + \alpha_e < \pi$, $\varepsilon > 0$ and for $p$ verifying (10) there exist a convex-concave meniscus on $[R_a - \varepsilon, R_a]$ then there exists $\varepsilon_1$ such that : $0 < \varepsilon_1 < \varepsilon$, the convex-concave meniscus on $[R_a - \varepsilon_1, R_a]$ is a concave meniscus and the convex- concave meniscus is static unstable.

## NUMERICAL RESULTS AND COMMENTS

The values of parameters used in numerical computation for dewetted InSb growth are the followings: $\gamma = 0.42[N \cdot m^{-1}]$; $\rho = 6582[kg \cdot m^{-3}]$; $\theta_c = 1.953[rad]$; $\alpha_e = 0.436[rad]$; $H_m = 60 \cdot 10^{-3}[m]$; $R_a = 5.5 \cdot 10^{-3}[m]$; $g = 9.81[m \cdot s^{-2}]$ and $\Delta P = P_{cold} - P_{hot} = \rho \cdot g \cdot H_m - p$.

**i). Static stability and instability ranges in case of meniscus with concave meridian curve.**

Using Statement 4.formula (10) is found that :if a concave meniscus having a gap size $\varepsilon$ in the range $(0, R_a)$ exists ,then for $\varepsilon$ in the range $(0, 1.802888752 \cdot 10^{-3})[m]$ the meniscus is static stable and for $\varepsilon$ in the range $(4.865294191 \cdot 10^{-3}, 5.5 \cdot 10^{-3})[m]$ the meniscus is static unstable. See Fig.2.

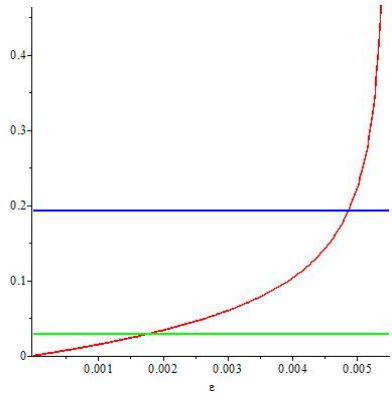

Fig.2.Static stability and instability ranges in case of InSb meniscus with concave meridian curve.

Remark that if the gap size is in the range $(1.802888752 \cdot 10^{-3}[m], 4.865294191 \cdot 10^{-3}[m])$ no information concerning the static stability or instability of the meniscus having concave meridian curve (in case when the meniscus exists).

**ii.) Existence of static -stable meniscus with concave meridian curve**.
Using Statement2 is found that: if the pressure difference $\Delta P$ verifies inequality
$\Delta P = P_{cold} - P_{hot} = \rho \cdot g \cdot H_m - p > 3845.6839 - l(1.802888752 \cdot 10^{-3}) = 4229.7379[Pa]$,
then for that pressure difference a static- stable meniscus having concave meridian curve is obtained, and the gape size $\varepsilon$ of the obtained meniscus is in the range $(0, 1.802888752 \cdot 10^{-3})[m]$.

**iii). Existence of a range of gap sizes for which static-stable meniscus with concave meridian curve exist**.
Because condition (8) from Statement3, concerning the pressure difference, is verified for any $\varepsilon$ verifying $1 \cdot 10^{-5} < \varepsilon < 1.831 \cdot 10^{-5}[m]$ it follows that for any gape size in the range

$(1 \cdot 10^{-5}, 1.831 \cdot 10^{-5})[m]$, static- stable meniscus having concave meridian curve exist. (see Fig.3).

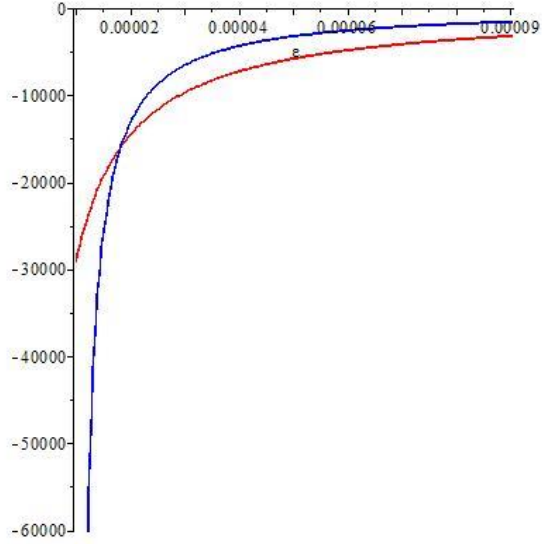

Fig.3 Existence of concave meridian curve for $1 \cdot 10^{-5} < \varepsilon < 1.831 \cdot 10^{-5}[m]$.

For instance if $\varepsilon = 1.5 \cdot 10^{-5}[m]$ then the concave meridian curve is obtained for the pressure difference $\Delta P = 18734.1652[Pa]$ and the meniscus height is $h_c = 1.421 \cdot 10^{-5}[m]$.

From the point of view of static - stability the seeding can be successful in case of menisci having concave meridian curve and gape size in the range $(1 \cdot 10^{-5}, 1.831 \cdot 10^{-5})[m]$.

**iv). Existence of static -stable meniscus with concave meridian curve and gap size in the range** $[1.831 \cdot 10^{-5}, 1.802888752 \cdot 10^{-3})[m]$

The effective determination of a concave meridian curve, for a given gape size in the range $[1.831 \cdot 10^{-5}, 1.802888752 \cdot 10^{-3})[m]$ (if exists) can be made determining the corresponding pressure difference limits, using formula (6) Statement1, and integrating numerically the initial value problem:

$$\begin{cases} \dfrac{dz}{dr} = \tan\theta \\ \dfrac{d\theta}{dr} = \dfrac{-\rho \cdot g \cdot z + \rho \cdot g \cdot H_m - \Delta P}{\gamma} \cdot \dfrac{1}{\cos\theta} - \dfrac{1}{r} \cdot \tan\theta \\ z(R_a) = 0, \quad \theta(R_a) = \theta_c - \dfrac{\pi}{2} \end{cases}$$

(11)

for different values of $\Delta P$ in the obtained pressure difference range.

The limits of the pressure difference $\Delta P$, computed according to formula (6), are represented on Fig.4.

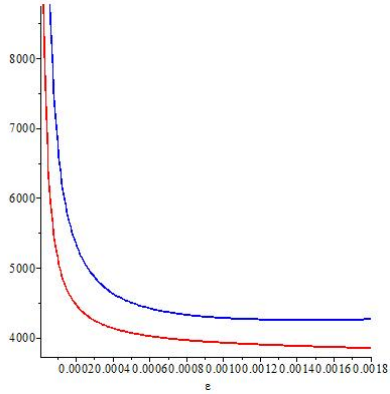 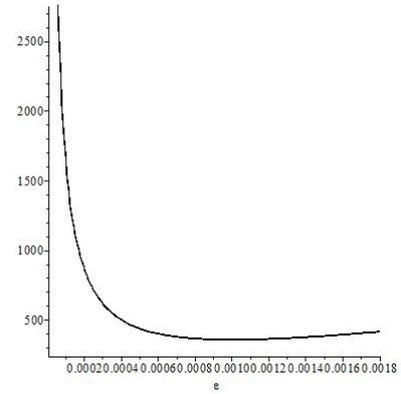

Fig.4a. Limits of the pressure differences $\Delta P$ for gap size in the range $[1.831 \cdot 10^{-5}, 1.802888752 \cdot 10^{-3})[m]$.

Fig.4b The breadth of the pressure difference ranges versus the gap size

The gap sizes $\varepsilon_1 = 5 \cdot 10^{-4}[m]$, $\varepsilon_2 = 1 \cdot 10^{-3}[m]$, $\varepsilon_3 = 1.802 \cdot 10^{-3}[m]$ belong to the above range. Computing the pressure limits, for these gap sizes the following ranges were obtained:
$[l(\varepsilon_1), L(\varepsilon_1)] = [4036.521, 4473.060][Pa]$; $[l(\varepsilon_2), L(\varepsilon_2)] = [3894.571, 4249.074][Pa]$; $[l(\varepsilon_3), L(\varepsilon_3)] = [3816.812, 4229.694][Pa]$.

In order to find the appropriate pressure differences $\Delta P = P_{cold} - P_{hot}$, the problem (11) was solved numerically for different values of $\Delta P$ in the computed ranges. In this way it was found that the values $(\Delta P)_1, (\Delta P)_2, (\Delta P)_3$ for which the gap sizes are $\varepsilon_1 = 5 \cdot 10^{-4}[m]$, $\varepsilon_2 = 1 \cdot 10^{-3}$, $\varepsilon_3 = 1.802 \cdot 10^{-3}$ and $\theta(R_c) = \pi/2 - \alpha_e = 1.134\ [rad]$ are the followings:
$(\Delta P)_1 = 4281.5652[Pa]$, $(\Delta P)_2 = 4065.8652[Pa]$ and $(\Delta P)_3 = 3978.8652[Pa]$ respectively. The shape and size of the computed menisci are represented on Fig.5-Fig.7.

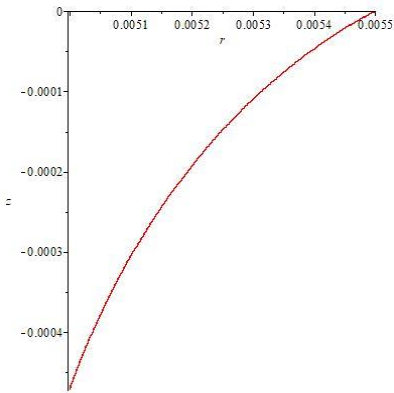 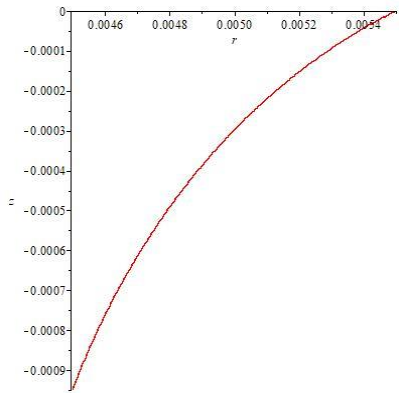

Fig.5 $z = z(r)$; $(\Delta P)_1 = 4281.5652[Pa]$
$\varepsilon_1 = 5 \cdot 10^{-4}[m]$; $R_c = 5 \cdot 10^{-3}[m]$;
$h_c = 4.718 \cdot 10^{-4}[m]$

Fig.6 $z = z(r)$; $(\Delta P)_2 = 4065.8652[Pa]$
$\varepsilon_2 = 1 \cdot 10^{-3}[m]$; $R_c = 4.5 \cdot 10^{-3}[m]$
$h_c = 9.501 \cdot 10^{-4}[m]$

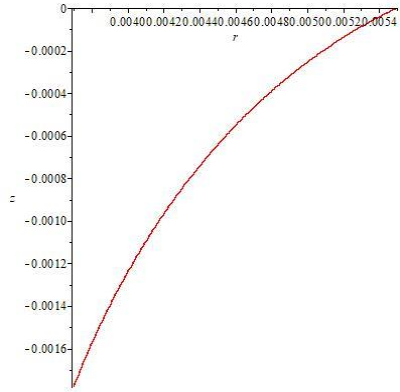

Fig.7 $z = z(r)$; $(\Delta P)_3 = 3978.8652[Pa]$
$\varepsilon_3 = 1.802 \cdot 10^{-3}[m]$; $R_c = 3.698 \cdot 10^{-3}[m]$;
$h_c = 1.78 \cdot 10^{-3}[m]$

It follows that the seeding can be successful in the cases:
gap size $\varepsilon_1 = 5 \cdot 10^{-4}[m]$; $(\Delta P)_1 = 4281.5652[Pa]$; seed radius equal to $R_c = 5 \cdot 10^{-3}[m]$; crystallization front height equal to $h_c = 4.718 \cdot 10^{-4}[m]$.

gap size $\varepsilon_2 = 1 \cdot 10^{-3}[m]$; $(\Delta P)_2 = 4065.8652[Pa]$; seed radius equal to $R_c = 4.5 \cdot 10^{-3}[m]$; crystallization front height equal to $h_c = 9.501 \cdot 10^{-4}[m]$.

gap size $\varepsilon_3 = 1.802 \cdot 10^{-3}[m]$; $(\Delta P)_3 = 3978.8652[Pa]$; seed radius equal to $R_c = 3.698 \cdot 10^{-3}[m]$; crystallization front height equal to $h_c = 1.78 \cdot 10^{-3}[m]$.

Comparing the above results it is interesting to observe that for a relatively narrow range of the pressure difference $\Delta P$ i.e. $[18734.1652, 3978.8652][Pa]$ it is possible to obtain a relatively large gap size range i.e. $[1.5 \cdot 10^{-5}, 1.802 \cdot 10^{-3}][m]$. It is also interesting to remark that as the gap size approaches to the upper limit $1.802888752 \cdot 10^{-3}[m]$ of the range of gap sizes, for which we have static stability, the meniscus height increases approaching to the value of $h_c = 1.78 \cdot 10^{-3}[m]$. This relatively high meniscus can be an explanation of the dynamic instability. But from the point of view of the static stability the above menisci are static stables and are appropriate for seeding.

**v). Existence of meniscus with convex- concave meridian curve**.
In [7] for $\varepsilon = 2.2 \cdot 10^{-3}[m]$ and $\Delta P = 4005[Pa]$ a convex- concave meniscus of height $h_c = 6.1 \cdot 10^{-3}[m]$ was found. This show that there is a change when the gap size $\varepsilon$ (we want to obtain) is more than $1.802888752 \cdot 10^{-3}[m]$. For instance if $\varepsilon = 2.143 \cdot 10^{-3}[m]$ and $\Delta P = 4005.662[Pa]$ then according to our computation a convex-concave meniscus is obtained as is shown on Fig.8

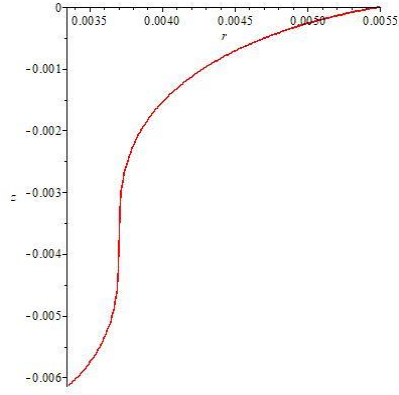

Fig.8 $z = z(r)$; $\Delta P = 4005.662 [Pa]$;
$\varepsilon = 2.143 \cdot 10^{-3} [m]$; $R_c = 3.357 \cdot 10^{-3} [m]$;
$h_c = 6.1388 \cdot 10^{-3} [m]$;

;

According to Statement 6 this meniscus is static-unstable. So it is inappropriate for seeding. Statement 5 and 6 concern general conditions for the existence and property of static meniscus with convex-concave meridian curve.

For the gap size in the range $(1.803 \cdot 10^{-3}, 5 \cdot 10^{-3}) [m]$, the computed pressure difference limits $\Delta P$ versus gap size for convex-concave meridian curve are presented in Fig.9.

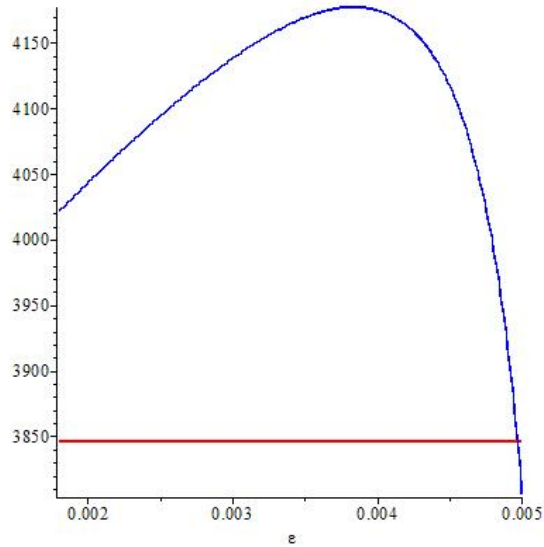

Fig.9. Pressure difference limits $\Delta P$ versus gap size for convex-concave meridian curve.

For the gap sizes $\varepsilon_4 = 2.15 \cdot 10^{-3} [m]$; $\varepsilon_5 = 2.2 \cdot 10^{-3} [m]$; $\varepsilon_6 = 2.5 \cdot 10^{-3} [m]$; $\varepsilon_7 = 3 \cdot 10^{-3} [m]$; the computed pressure difference limits are: $l_4 = 3845.683 [Pa]$, $L_4 = 4143.652 [Pa]$; $l_5 = 3845.683 [Pa]$, $L_5 = 4150.581 [Pa]$; $l_6 = 3845.683 [Pa]$, $L_6 = 4192.158 [Pa]$; $l_7 = 3845.683 [Pa]$, $L_7 = 4261.453 [Pa]$.

The appropriate pressure differences for which the meridian curves of these menisci are convex-concave are: $(\Delta P)_4 = 4005.4552 [Pa]$; $(\Delta P)_5 = 4003.765 [Pa]$; $(\Delta P)_6 = 3993.265 [Pa]$; $(\Delta P)_7 = 3973.4652 [Pa]$.

The height of the menisci for the gap sizes $\varepsilon_4$, $\varepsilon_5$, $\varepsilon_6$, $\varepsilon_7$ are $h_c^4 = 6.135 \cdot 10^{-3}[m]$; $h_c^5 = 6.09 \cdot 10^{-3}[m]$; $h_c^6 = 5.82 \cdot 10^{-3}[m]$; $h_c^7 = 5.26 \cdot 10^{-3}[m]$; and on every meridian curve the growth angle is achieved twice as follows : $R^4{}_c = 3.35 \cdot 10^{-3}$ , $r^4 = 4.07 \cdot 10^{-3}$ ; $R^5{}_c = 3.3 \cdot 10^{-3}$, $r^5 = 4.052 \cdot 10^{-3}$ ; $R^6{}_c = 3 \cdot 10^{-3}$ , $r^6 = 3.925 \cdot 10^{-3}$ ; $R^7{}_c = 2.5 \cdot 10^{-3}$, $r^7 = 3.587 \cdot 10^{-3}$.

The menisci with convex-concave meridian curves obtained for $(\Delta P)_4 = 4005.4552[Pa]$; $(\Delta P)_5 = 4003.765[Pa]$; $(\Delta P)_6 = 3993.265[Pa]$ ; $(\Delta P)_7 = 3973.4652[Pa]$ are static unstable and inappropriate for seeding.

The gap sizes of the concave part of the meridian curves of menisci obtained for the same values of the pressure differences are: $\varepsilon'_4 = 1.43 \cdot 10^{-3}[m]$; $\varepsilon'_5 = 1.448 \cdot 10^{-3}[m]$; $\varepsilon'_6 = 1.575 \cdot 10^{-3}[m]$; $\varepsilon'_7 = 1.913 \cdot 10^{-3}[m]$;

The gap sizes $\varepsilon'_4$, $\varepsilon'_5$, $\varepsilon'_6$ are in the range $[5 \cdot 10^{-4}, 1.802 \cdot 10^{-3}][m]$ and so the menisci having concave meridian curves and corresponding to the pressure differences $(\Delta P)_4 = 4005.4552[Pa]$; $(\Delta P)_5 = 4003.765[Pa]$; $(\Delta P)_6 = 3993.265[Pa]$ ; are static stable.

The height of the above static-stable menisci having concave meridian curves are: $h^4 = 1.38 \cdot 10^{-3}[m]$; $h^5 = 1.40 \cdot 10^{-3}[m]$; $h^6 = 1.53 \cdot 10^{-3}[m]$;.

It follows that the seeding can be successful in the cases:

$(\Delta P)_4 = 4005.4552[Pa]$, seed radius equal to $r^4 = 4.07 \cdot 10^{-3}[m]$ ,crystallization front height equal to $h^4 = 1.38 \cdot 10^{-3}[m]$; gap size : $\varepsilon'_4 = 1.43 \cdot 10^{-3}[m]$;

$(\Delta P)_5 = 4003.765[Pa]$, seed radius equal to $r^5 = 4.052 \cdot 10^{-3}[m]$ ,crystallization front height equal to $h^5 = 1.40 \cdot 10^{-3}[m]$; gap size $\varepsilon'_5 = 1.448 \cdot 10^{-3}[m]$;

$(\Delta P)_6 = 3393.265[Pa]$, seed radius equal to $r^6 = 3.925 \cdot 10^{-3}[m]$ ,crystallization front height equal to ape size $h^6 = 1.53 \cdot 10^{-3}[m]$; gap size $\varepsilon'_6 = 1.575 \cdot 10^{-3}[m]$;

The height of the static meniscus having concave meridian curve corresponding to the pressure difference $(\Delta P)_7 = 3973.4652[Pa]$ is $h^7 = 1.90 \cdot 10^{-3}[m]$; but its gap size is $\varepsilon'_7 = 1.913 \cdot 10^{-3}[m]$. Because the gap size is not anymore in the range $[1.831 \cdot 10^{-5}, 1.802888752 \cdot 10^{-3})[m]$ Statement 4 concerning static stability /instability can't be applied.

It is interesting to observe that for the gap size $\varepsilon_8 = 3.5 \cdot 10^{-3}[m]$ the pressure difference limits are $l_8 = 3845.683[Pa]$ , $L_8 = 4330.784[Pa]$ but for $\Delta P$ in the range $(l_8, L_8)$ we don't find convex-concave meridian curve.

## CONCLUSIONS

1.The six theoretical results provide information concerning the existence, stability, instability of concave meniscus as well the existence of convex-concave menisci in terms of the gap size. This information's are new and can help the seed size choose as well the thermal conditions preparations for a dewetted Bridgman process.

2.The numerical results show that in case of InSb semiconductor the theoretical results are effective and from the static stability point of view reveal situations when the seeding can be successful and situations when the seeding is not successful.

# APPENDIX.

**Proof of the Statement1.** Let $\theta_c + \alpha_e < \pi$, $R_c = R_a - \varepsilon$, and $z(r)$ defined for $r \in [R_c, R_a]$ which verifies (1)-(4) and $z''(r) < 0$. The function defined as:

$$\theta(r) = \arctan z'(r) \quad \text{for } r \in [R_c, R_a]$$

verifies

$$\theta'(r) = \frac{-\rho \cdot g \cdot z(r) + p}{\gamma} \cdot \frac{1}{\cos(\theta(r))} - \frac{1}{r} \cdot \tan(\theta(r))$$

and the boundary conditions: $\theta(R_a) = \theta_c - \pi/2$, $\theta(R_c) = \pi/2 - \alpha_e$.

Hence, by the mean value theorem, there exists $r' \in [R_c, R_a]$ such that the following equality holds:

$$p = \gamma \cdot \frac{\theta_c + \alpha_e - \pi}{R_a - R_c} \cdot \cos\theta(r') + \rho \cdot g \cdot z(r') + \frac{\gamma}{r'} \cdot \sin(\theta(r'))$$

On the other hand, inequality $z''(r) < 0$ implies that the function $z'(r)$ is strictly decreasing and by consequence the function $\theta(r)$ is strictly decreasing. Therefore, the following inequalities hold:

$\theta_c - \pi/2 < \theta(r') < \pi/2 - \alpha_e$; $\sin\alpha_e \le \cos\theta(r') \le \sin\theta_c$; $-\cos\theta_c \le \sin\theta(r') \le \cos\alpha_e$;

$$-\rho \cdot g \cdot \varepsilon \cdot \tan(\frac{\pi}{2} - \alpha_e) \le \rho \cdot g \cdot z(r') \le -\rho \cdot g \cdot (R_a - r') \cdot \tan(\theta_c - \frac{\pi}{2}) \le 0.$$

Hence inequality (6) is obtained.

**Proof of the Statement2.** Consider the function $z(r)$ which verify (1), (3). Denote by I the maximal interval on which the function $z(r)$ exists and by $\theta(r)$ the function defined by $\theta(r) = \arctan z'(r)$. This function verify $\theta'(r) = \frac{-\rho \cdot g \cdot z(r) + p}{\gamma} \cdot \frac{1}{\cos(\theta(r))} - \frac{1}{r} \cdot \tan(\theta(r))$. Because

$$z''(R_a) = \frac{\theta'(R_a)}{\cos^2\theta(R_a)} = \frac{1}{\cos^3\theta(R_a)} \cdot [\frac{p}{\gamma} - \frac{\sin\theta(R_a)}{R_a}] <$$

$$\frac{1}{\cos^3\theta(R_a)}[\frac{\theta_c + \alpha_e - \pi}{\varepsilon'} \cdot \sin\theta_c - \frac{1}{\gamma}\rho \cdot g \cdot \varepsilon' \cdot \tan(\frac{\pi}{2} - \alpha_e)] < 0;$$

$z'(R_a) = \tan(\theta_c - \frac{\pi}{2}) > 0$ and $z'(R_a) < \tan(\frac{\pi}{2} - \alpha_e)$;

there exists $r' \in I, 0 < r' < R_a$ such that for any $r$ which verify $r' \le r \le R_a$ the following inequalities hold: $z''(r) < 0$; $z'(r) > \tan(\theta_c - \frac{\pi}{2})$ and $z'(r) < \tan(\frac{\pi}{2} - \alpha_e)$. Let $r_*$ be the infimum of the set of numbers $r'$ for which the above conditions hold, $r_* = \inf\{r'\}$.

Remark that for $r_* + 0$ the following inequalities hold: $z(r_* + 0) \le z(r) \le 0$ for any $r$ which verify $r_* \le r < R_a$, $-(R_a - r_*) \cdot \tan(\frac{\pi}{2} - \alpha_e) \le z(r_* + 0) \le -(R_a - r_*) \cdot \tan(\theta_c - \frac{\pi}{2})$

Now we will show that $\varepsilon = (R_a - r_*) \le \varepsilon'$ and $z'(r_* + 0) = \tan(\frac{\pi}{2} - \alpha_e)$.

For showing inequality $\varepsilon = (R_a - r_*) \leq \varepsilon'$ assume the contrary i.e. $\varepsilon = (R_a - r_*) > \varepsilon'$. Under this hypothesis for some $r' \in (R_a - \varepsilon, R_a)$ the following relations hold:

$$\theta(R_a - \varepsilon') - \theta(R_a) = -\theta'(r') \cdot \varepsilon' = \frac{1}{\cos\theta(r')} \cdot [-\frac{p}{\gamma} + \frac{\rho \cdot g \cdot z(r')}{\gamma} + \frac{1}{r'} \cdot \sin\theta(r')] \cdot \varepsilon' >$$

$$\varepsilon' \cdot \frac{1}{\cos\theta(r')} \cdot [-\frac{\theta_c + \alpha_e - \pi}{\varepsilon'} \cdot \sin\theta_c + \frac{1}{\gamma}\rho \cdot g \cdot \varepsilon' \cdot \tan\left(\frac{\pi}{2} - \alpha_e\right) + \frac{1}{R_a} \cdot \cos\theta_c + \frac{\rho \cdot z(r')}{\gamma} + \frac{1}{r'} \cdot \sin\theta(r')]$$

$> \pi - \theta_c - \alpha_e$. Hence $\theta(R_a - \varepsilon') > \frac{\pi}{2} - \alpha_e$ what is impossible according to the definition of $r_*$.

In order to show that $z'(r_* + 0) = \tan(\frac{\pi}{2} - \alpha_e)$ we remark that from the definition of $r_*$ it follows that in $r_*$ one of the following three equalities hold: $z''(r_* + 0) = 0$ or $z'(r_* + 0) = \tan(\theta_c - \frac{\pi}{2})$, or $z'(r_* + 0) = \tan(\frac{\pi}{2} - \alpha_e)$.

Since $z'(r_* + 0) \leq z'(r) < \tan(\theta_c - \frac{\pi}{2})$ for $r \in (r_* + 0, R_a)$ equality $z'(r_* + 0) = \tan(\theta_c - \frac{\pi}{2})$ is impossible. It follows that at $r_*$ only one of the following two equalities hold $z''(r_* + 0) = 0$ or $z'(r_* + 0) = \tan(\frac{\pi}{2} - \alpha_e)$.

Now we will show that the equality $z''(r_* + 0) = 0$ is impossible. For that assume the contrary, that is $z''(r_* + 0) = 0$. Hence:

$$p = \rho \cdot g \cdot z(r_* + 0) + \frac{\gamma}{r_*} \cdot \sin\theta(r_* + 0) \geq -\rho \cdot g \cdot (R_a - r_*) \cdot \tan(\frac{\pi}{2} - \alpha_e) + \frac{\gamma}{r_*} \cdot \sin\theta(r_* + 0) >$$

$$-\rho \cdot g \cdot \varepsilon' \cdot \tan(\frac{\pi}{2} - \alpha_e) + \frac{\gamma}{R_a} \cdot \sin(\theta_c - \frac{\pi}{2}) >$$

$$\gamma \cdot \frac{\theta_c + \alpha_e - \pi}{\varepsilon'} \cdot \sin\theta_c - \rho \cdot g \cdot \varepsilon' \cdot \tan\left(\frac{\pi}{2} - \alpha_e\right) - \frac{\gamma}{R_a} \cdot \cos\theta_c$$

and that is impossible. Therefore in $r_*$ equality $z'(r_* + 0) = \tan(\frac{\pi}{2} - \alpha_e)$ holds.

**Proof of the Statement 3.** According to the Statement 2 for any $p$ which verify inequality $p < L(\varepsilon_2)$ there exists a number $\varepsilon$ having the property $0 < \varepsilon < \varepsilon_2$ and function $z(r)$ having the properties (1) - (4) and $z''(r) < 0$ for $r \in [R_c, R_a]$ with $R_c = R_a - \varepsilon$. Assume now that $p$ verify also the inequality $l(\varepsilon_1) < p$ and show that $\varepsilon_1 < \varepsilon$. Assuming the contrary i.e. $\varepsilon < \varepsilon_1$ it is easy to show that the following inequality hold :

$\gamma \cdot \frac{\theta_c + \alpha_e - \pi}{\varepsilon} \cdot \sin\alpha_e + \frac{\gamma}{R_a - \varepsilon} \cdot \cos\alpha_e < l(\varepsilon_1)$. On the other hand according to the Statement 1 for $p$ the following inequality hold $p \leq \gamma \cdot \frac{\theta_c + \alpha_e - \pi}{\varepsilon} \cdot \sin\alpha_e + \frac{\gamma}{R_a - \varepsilon} \cdot \cos\alpha_e$. Hence

$p < l(\varepsilon_1)$ what is in contradiction with the left hand side of the inequality (9) appearing in the hypothesis.

**Proof of the Statement 4.**

Since (1) is the Euler equation ( [9] Chapter2) for the free energy functional (5), in this case it is sufficient to investigate the Legendre and Jacobi conditions ( [9] Chapter 8). Consider for that the function

$$F(z,z',r) = \left\{ \gamma \cdot [1+(z')^2]^{1/2} - \frac{1}{2} \cdot \rho \cdot g \cdot z^2 + p \cdot z \right\} \cdot r$$

and remark that the Legendre condition $\frac{\partial^2 F}{\partial z'^2} > 0$ reduces to the inequality:

$$r \cdot \gamma [1+(z')^2]^{-3/2} > 0$$

which is verified.

The Jacobi equation

$$[\frac{\partial^2 F}{\partial z^2} - \frac{d}{dr}(\frac{\partial^2 F}{\partial z \partial z'})] \cdot \eta - \frac{d}{dr}[\frac{\partial^2 F}{\partial z'^2} \cdot \eta'] = 0$$

in this case become

$$\frac{d}{dr}(\frac{r \cdot \gamma}{[1+(z')^2]^{3/2}} \cdot \eta') + \rho \cdot g \cdot r \cdot \eta = 0$$

i). For obtaining the stability result remark that for the coefficients of the Jacobi equation the following inequalities hold:

$$\frac{r \cdot \gamma}{[1+(z')^2]^{3/2}} \geq (R_a - \varepsilon) \cdot \gamma \cdot \sin^3 \alpha_e \qquad\qquad \rho \cdot g \cdot r \leq \rho \cdot g \cdot R_a$$

Therefore, the equation

$$\frac{d}{dr}((R_a - \varepsilon) \cdot \gamma \cdot \sin^3 \alpha_e \cdot \varsigma') + \rho \cdot g \cdot R_a \cdot \varsigma = 0$$

is a „Sturm type upper bound" ( [9] Chapter 11) for the Jacobi equation . An arbitrary solution of the above „Sturm type upper bound equation" is given by

$$\varsigma(r) = A \cdot \sin(\omega \cdot r + \varphi)$$

Here $A$ and $\varphi$ are arbitrary real constants and $\omega^2 = \frac{\rho \cdot g}{\gamma \cdot \sin^3 \alpha_e} \cdot \frac{R_a}{R_a - \varepsilon}$. The half period of any non-zero solution $\varsigma(r)$ is $\frac{\pi}{\omega} = \pi \cdot \frac{\gamma^{1/2} \cdot \sin^{3/2} \alpha_e}{\rho^{1/2} \cdot g^{1/2}} \cdot \frac{(R_a - \varepsilon)^{1/2}}{R_a^{1/2}}$. If the half period is more than $R_a - R_c = \varepsilon$ then any non-zero solution $\varsigma(r)$ vanishes at most once on the interval $[R_c, R_a]$ .In other words if the following inequality hold

$$\pi \cdot \frac{\gamma^{1/2} \cdot \sin^{3/2} \alpha_e}{\rho^{1/2} \cdot g^{1/2}} \cdot \frac{(R_a - \varepsilon)^{1/2}}{R_a^{1/2}} > \varepsilon \quad \text{or} \quad \frac{\varepsilon}{(R_a - \varepsilon)^{1/2}} < \pi \cdot \frac{1}{R_a^{1/2}} \cdot \frac{\gamma^{1/2} \cdot \sin^{3/2} \alpha_e}{\rho^{1/2} \cdot g^{1/2}}$$

then any non-zero solution $\varsigma(r)$ vanishes at most once on the interval $[R_c, R_a]$. Hence, according to [9] Chapter 11, the solution $\eta(r)$ of Jacobi equation which verifies $\eta(R_a) = 0$ and $\eta'(R_a) = 1$, has only one zero on the interval $[R_c, R_a]$. This means that the Jacobi condition for weak minimum is verified [9].

ii). For obtaining the instability result remark that for the coefficients of the Jacobi equation the following inequalities hold:

$$\frac{r \cdot \gamma}{[1 + (z')^2]^{3/2}} \leq R_a \cdot \gamma \cdot \cos^3(\theta_c - \frac{\pi}{2}) \quad \text{and} \quad \rho \cdot g \cdot r \geq \rho \cdot g \cdot (R_a - \varepsilon).$$

Therefore, the equation

$$\frac{d}{dr}(R_a \cdot \gamma \cdot \sin^3 \theta_c \cdot \xi') + \rho \cdot g \cdot (R_a - \varepsilon) \cdot \xi = 0$$

is a „Sturm type lower bound equation" ([9] Chap.11) for the Jacobi equation. An arbitrary solution of the above „Sturm type lower bound equation" is given by

$$\xi(r) = A \cdot \sin(\omega \cdot r + \varphi)$$

Here $A$ and $\varphi$ are arbitrary real constants and $\omega^2 = \frac{\rho \cdot g \cdot (R_a - \varepsilon)}{R_a \cdot \gamma \cdot \sin^3 \theta_c}$. The period of any non-zero solution $\xi(r)$ is $\frac{2 \cdot \pi}{\omega} = 2 \cdot \pi \cdot \frac{\gamma^{1/2} \cdot \sin^{3/2} \theta_c}{\rho^{1/2} \cdot g^{1/2}} \cdot \frac{(R_a - \varepsilon)^{1/2}}{R_a^{1/2}}$. If the period is less than $R_a - R_c$ then any non-zero solution $\xi(r)$ vanishes at least twice on the interval $[R_c, R_a]$. In other words if the following inequality hold :

$$2 \cdot \pi \cdot \frac{\gamma^{1/2} \cdot \sin^{3/2} \theta_c}{\rho^{1/2} \cdot g^{1/2}} \cdot \frac{(R_a - \varepsilon)^{1/2}}{R_a^{1/2}} < \varepsilon \quad \text{or} \quad \frac{\varepsilon}{(R_a - \varepsilon)^{1/2}} > 2 \cdot \pi \cdot \frac{1}{R_a^{1/2}} \cdot \frac{\gamma^{1/2} \cdot \sin^{3/2} \theta_c}{\rho^{1/2} \cdot g^{1/2}}$$

then any non-zero solution $\xi(r)$ vanishes at least twice on the interval $[R_c, R_a]$. Hence, according to [9] Chapter 11, that solution $\eta(r)$ of Jacobi equation which satisfies $\eta(R_a) = 0$ and $\eta'(R_a) = 1$ vanishes at least twice on the interval $[R_c, R_a]$. This means that the Jacobi condition for weak minimum is not satisfied [9].

**Proof of the Statement 5.** Using equation (1) it is easy to see that condition $z''(R_a) < 0$ implies inequality $-\frac{\gamma}{R_a} \cdot \cos\theta_c > p$. For the left hand side of (10) remark that condition $z''(R_c) > 0$ implies that $p > \frac{\gamma}{R_a - \varepsilon} \cdot \cos\alpha_e - \rho \cdot g \cdot \varepsilon \cdot \tan(\frac{\pi}{2} - \alpha_e)$

**Proof of the Statement 6.** Since $z''(R_c) > 0$ it follows that $\theta(r) > \theta(R_a - \varepsilon)$ for $r > R_a - \varepsilon$, $r$ sufficiently close to $R_a - \varepsilon$ i.e. $\theta(r) > \frac{\pi}{2} - \alpha_e$. Due to the fact that $\theta(R_a) = \theta_c - \frac{\pi}{2} < \frac{\pi}{2} - \alpha_e$ it follows that there exist $r_* \in (R_a - \varepsilon, R_a)$ such that $\theta(r_*) = \frac{\pi}{2} - \alpha_e$. So for $\varepsilon_1 = R_a - r_*$ we have $\theta(R_a - \varepsilon_1) = \frac{\pi}{2} - \alpha_e$.

The static instability of the convex-concave meniscus is a consequence of the inequality:

$$\int_{R_c}^{R_a} \left\{ \gamma \cdot [1 + (z')^2]^{1/2} - \frac{1}{2} \cdot \rho \cdot g \cdot z^2 + p \cdot z \right\} \cdot r \cdot dr > \int_{r_*}^{R_a} \left\{ \gamma \cdot [1 + (z')^2]^{1/2} - \frac{1}{2} \cdot \rho \cdot g \cdot z^2 + p \cdot z \right\} \cdot r \cdot dr$$

**ACKNOWLEDGMENT**

This research did not receive any specific grant from funding agencies in the public, commercial or not-for-profit sectors.